\documentclass{appolb}
\usepackage{graphicx}
\usepackage{slashed}
\usepackage{amsmath}  
\usepackage{epstopdf}
\usepackage{float}
\usepackage[dvipsnames]{xcolor}

\begin{document}
\title{Heavy and heavy-light mesons and the Lorentz structure of the quark-antiquark interaction%
\thanks{Presented by A.\ Stadler at Excited QCD 2017, Sintra, Portugal}%
}

\author{S. Leit\~ao$^{1}$, A. Stadler$^{2,1}$,  M. T. Pe\~na$^{3,1}$,  E. P. Biernat$^{1}$
\address{$^1$Centro de F\'isica Te\'orica de Part\'iculas (CFTP), Instituto Superior T\'ecnico, Universidade de Lisboa, 1049-001 Lisboa, Portugal\\
$^2$Departamento de F\'isica, Universidade de \'Evora, 7000-671 \'Evora, Portugal\\
$^3$Departamento de F\'isica, Instituto Superior T\'ecnico (IST), Universidade de Lisboa, 1049-001 Lisboa, Portugal
}
}
\maketitle
\begin{abstract}
We solve a Minkowski-space integral equation, derived in the Covariant Spectator Theory, for quark-antiquark bound states describing heavy and heavy-light mesons. The equation's kernel contains a one-gluon exchange interaction and a covariant generalization of a linear confining potential with a mixed scalar, pseudoscalar, and vector Lorentz structure, characterized by a continuous mixing parameter. We investigate to what extent the Lorentz structure of the confining kernel can be determined by fitting the mixing parameter to the meson spectrum.
\end{abstract}
  
\section{Introduction}

Already several decades ago it has become clear that many mesons can be well described as bound states of a quark and an antiquark, interacting through a linear and a one-gluon exchange (OGE) potential \cite{Eichten:1978tg,Godfrey:1985jk}. However, modern unified decriptions of all mesons have to go beyond these early works in several aspects, among which we emphasize a proper treatment of chiral symmetry and of relativity. Among the frameworks that meet these requirements figure prominently Lattice QCD (e.g., \cite{McNeile:2006,Burch:2006,Dudek:2008,Briceno:2017yq}) and the Dyson-Schwinger/Bethe-Salpeter (DS/BS) approach \cite{Burden:1997vl,Maris:1997,Fischer:2003yf,Eichmann:2009rw,Hilger:2015ty,Eichmann:2016bf}. 

We use the Covariant Spectator Theory (CST), which is related to the BS equation, but goes beyond ladder approximation by effectively taking contributions from crossed-ladder diagrams into account \cite{Gross:1969eb,Stadler:2011to}. Chiral symmetry can also be implemented exactly in CST \cite{Gross:1991te,Biernat:2014jt}, but in this work we calculate heavy and heavy-light mesons, where this aspect is of minor importance. Instead, we focus on the role of relativity,  and in particular on the Lorentz structure of the confining interaction, which is at present still an open problem. 

\section{The CST bound-state equation}

\begin{figure}[tbhp]
\begin{center}
\includegraphics[width=0.5\textwidth]{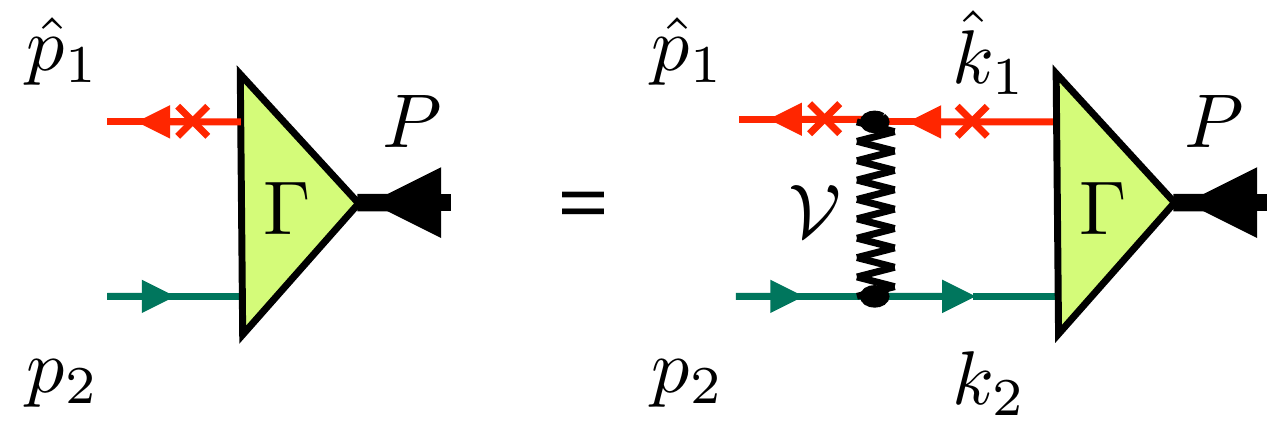}
\caption{The one-channel spectator equation (1CSE) for the bound-state vertex function $\Gamma$ of a quark (particle 1) and an antiquark (particle 2), interacting through a kernel $\cal{V}$. An ``$\times$'' on a line means that the particle is on its positive-energy mass shell, and the corresponding four-momentum carries a ``$\hat{\hspace{10pt}}$''.}
\label{fig:1CSE}
\end{center}
\end{figure}

The one-channel spectator equation (1CSE) for the meson-quark vertex function $\Gamma$ is shown graphically in Fig.~\ref{fig:1CSE}. It has the same structure as the Bethe-Salpeter equation, except that particle 1 (by convention the heavier particle, unless they have equal mass) is on its positive-energy mass shell. It is an approximation to the full, four-channel equation, which should work very well in systems with at least one heavy particle \cite{Leitao:2017mlx}.

We work in the meson's rest frame, where the total momentum is $P=\hat p_1-p_2=(\mu, {\bf 0})$,  $\hat{p}_1=(E_{1p},\bf{p})$ is the on-shell momentum of particle 1, with $E_{1p}=\sqrt{m_1^2+{\bf p}^2}$, 
and the external relative momentum is $p=(\hat p_1+p_2)/2=(E_{1p}-\mu/2,{\bf p})$  (analogous expressions hold for the internal relative momentum $k$). The 1CSE can be written \cite{Leitao:2017mlx}
\begin{multline}
  \Gamma(\hat p_1,p_2)  =
 - \int \frac{\mathrm d^3 k}{(2\pi)^3}\frac{m_1}{E_{1k}} \mathcal{V}(p,\hat{k}_1-P/2) \frac{m_1+\slashed{\hat k}_1}{2m_1} \Gamma(\hat{k}_1,k_2)  \frac{m_2+\slashed{k}_2}{m_2^2-k_2^2-i\epsilon} 
 \,,
 \label{eq:1CSE}
\end{multline}
where $k_2=\hat{k}_1-P$, and we use fixed quark masses $m_i$ (which will be replaced by dynamical mass functions in future work).
Our interaction kernel $\cal V$ has the form
\begin{multline}
{\cal V} (p,k)=
 \left[ (1-y) \left({\bf 1}_1\otimes {\bf 1}_2 + \gamma^5_1 \otimes \gamma^5_2 \right) - y\, \gamma^\mu_1 \otimes \gamma_{\mu 2} \right]V_\mathrm{L}(p,k) 
\\
-\gamma^\mu_1 \otimes \gamma_{\mu 2} \left[ V_\mathrm{OGE}(p,k)+V_\mathrm{C}(p,k) \right] \, ,
\label{eq:kernel}
\end{multline}
where
 $V_\mathrm{L}(p,k)$ is a covariant generalization of a linear confining potential, $V_\mathrm{OGE}(p,k)$ is the short-range one-gluon-exchange interaction in Feynman gauge, and $V_\mathrm{C}(p,k)$ a covariant form of a constant potential. We use a Pauli-Villars regularization with a cut-off parameter $\Lambda=2m_1$ for all mesons. The detailed form of the kernel is given in \cite{Leitao:2017mlx}.
  
The OGE and constant kernels are Lorentz-vector interactions. The Lorentz structure of the linear confining kernel in (\ref{eq:kernel}) is a mixture of equally-weighted scalar and pseudoscalar coupling (S+P) on one hand, which satisfies the axial-vector Ward-Takahashi identity \cite{Biernat:2014uq}, and vector coupling (V) on the other hand.
By adjusting the mixing parameter $y$, we can continuously vary the relative weight of these structures, with $y=0$ yielding a pure S+P coupling, and $y=1$ a pure V coupling. 
For any value of $y$, the nonrelativistic limit of this kernel is always, when transformed into coordinate space, the Cornell type potential $V(r)=\sigma r -\alpha_s/r -C$.  

For numerical calculations, Eq.~(\ref{eq:1CSE}) is converted into an eigenvalue equation for the so-called ``relativistic wave functions'', defined as spinor matrix elements of the vertex functions multiplied by the quark propagators. In the nonrelativistic limit they turn then into Schr\"odinger wave functions. By solving this eigenvalue problem we obtain, very conveniently, the masses and wave functions of a whole tower of excited states.

\section{Numerical results and discussion}

Working with the 1CSE, in \cite{Leitao:2017it} we performed least-square fits of the model parameters $\sigma$, $\alpha_s$, and $C$, to the masses of selected data sets containing heavy and heavy-light mesons with $J^P=0^\pm$ and $1^\pm$, while choosing fixed values for the quark masses and keeping $y=0$. In this work we allow $y$ and the quark masses to vary, which represents a considerable increase in required computing time. The data set S1 contains 9 meson states with $0^-$, the set S2 (S2$'$) contains 25 (24)  states with $0^\pm$ and $1^-$, and the set S3 contains 39 states with $0^\pm$ and $1^\pm$ (the states are listed in \cite{Leitao:2017mlx}). 


\begin{table*}[tbh]\small
\begin{center}
\begin{tabular}{  c  | c c c c cccc  }
  Model & $\sigma$  & $\alpha_s$ &$C$ & $y$ & $m_b$ & $m_c$  &$m_s$  &$m_q$ \\ \hline
  M0$_{\text{S1}}$       & 0.2493 & 0.3643 & 0.3491 & {\bf 0.0000} & {\bf 4.892}& {\bf 1.600} &{\bf 0.4478} & {\bf 0.3455} \\
    M1$_{\text{S1}}$& 0.2235 & 0.3941 & 0.0591  &0.0000 & 4.768 & 1.398& 0.2547& 0.1230 \\
    \hline
    M0$_{\text{S2}}$&  0.2247 & 0.3614 & 0.3377  & {\bf 0.0000} & {\bf 4.892} & {\bf 1.600} & {\bf 0.4478}& {\bf 0.3455}  \\
M1$_{\text{S2}}$&  0.1893 & 0.4126 & 0.1085  &0.2537 & 4.825 & 1.470& 0.2349& 0.1000 \\
\hline
M1$_{\text{S2}'}$& 0.2017 & 0.4013 & 0.1311 & 0.2677 &  4.822 & 1.464 & 0.2365 & 0.1000  \\
\hline
M1$_{\text{S3}}$& 0.2022 & 0.4129 & 0.2145  &0.2002 & 4.875 & 1.553 & 0.3679 & 0.2493  \\
M0$_{\text{S3}}$&  0.2058 & 0.4172 & 0.2821  &{\bf 0.0000}  & 4.917 & 1.624 & 0.4616 & 0.3514 
    \end{tabular}
\end{center}
    \caption{Kernel parameters of the different models considered in this work. Quark masses and $C$ are in GeV, $\sigma$ in GeV$^2$, $\alpha_s$ and $y$ are dimensionless. The values in boldface were held fixed during the fits of the respective models.}
    \label{tab:parameters}
    \end{table*}

\begin{figure}[htbp]
\begin{center}
    \includegraphics[width=0.97\textwidth]{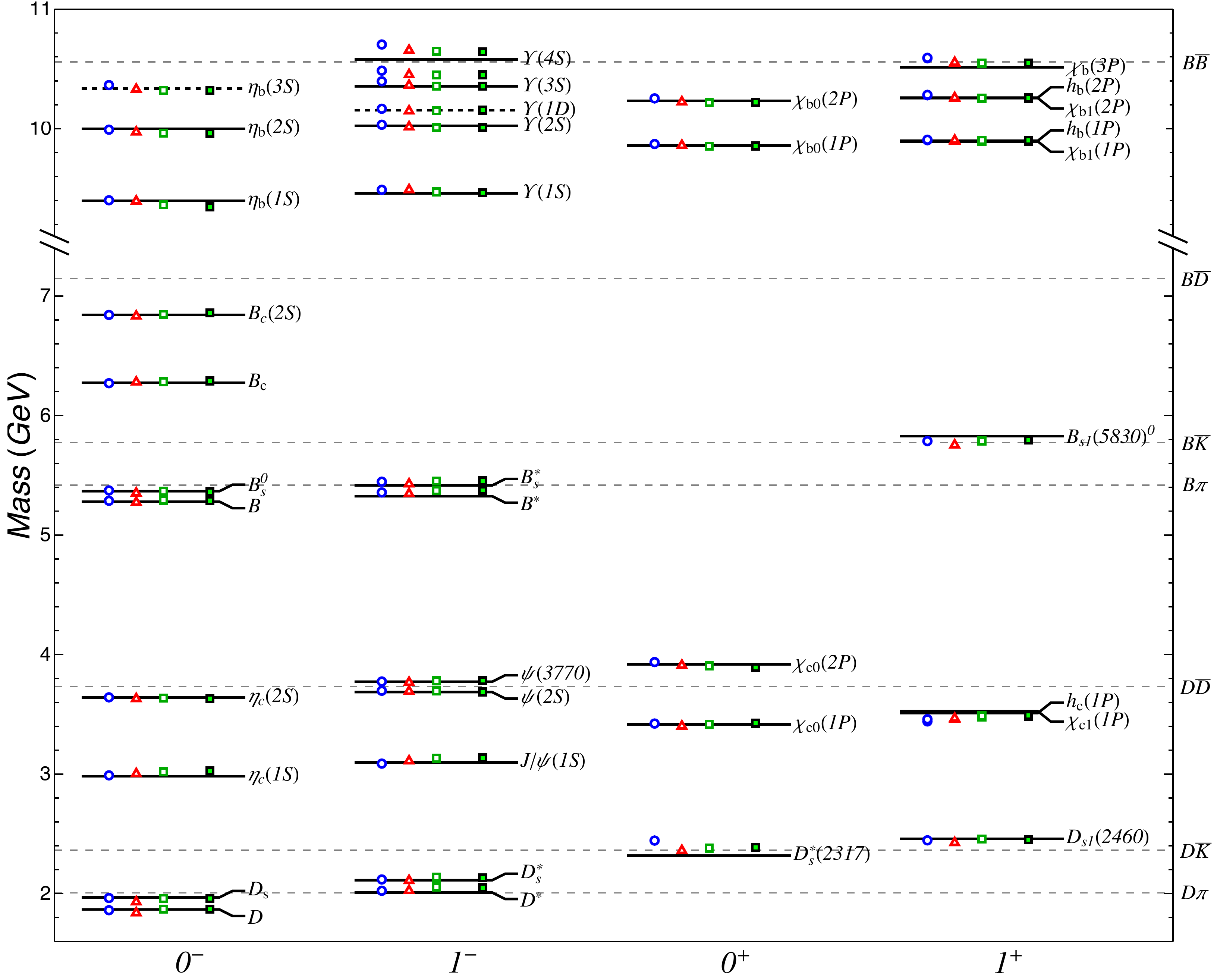}
\caption{Masses of heavy-light and heavy mesons with $J^P=0^\pm$ and $1^\pm$. The symbols represent the 1CSE results calculated with the models M1$_{\text{S1}}$ (circle), M1$_{\text{S2}'}$ (triangle), M1$_{\text{S3}}$ (open square), and M0$_{\text{S3}}$ (filled square) of Tab.~\ref{tab:parameters}.
Solid horizontal lines are the measured meson masses \cite{PDG2014}. Dashed horizontal lines across the figure indicate open flavor thresholds.}\label{fig:spectrum}
\end{center}
\end{figure}

Table \ref{tab:parameters} shows the parameters of the various models determined in our fits, which include the ones found in \cite{Leitao:2017it}, M0$_{\text{S1}}$ and M0$_{\text{S2}}$, for comparison. The results of four of our models are compared to the data in Fig.~\ref{fig:spectrum}. In fact, all models of Tab.~\ref{tab:parameters} reproduce the data very well, with rms differences to the largest data set S3 ranging from 30 to 40 MeV. This is a remarkable agreement, considering that we are performing \emph{global} fits, using the same model parameters for the whole range of mesons, with masses from below 2 GeV to over 10 GeV.

Fitting the quark masses does not lead to a significant improvement over the results of \cite{Leitao:2017it}. The light and charm quark masses can be varied by a few hundred MeV (the bottom quark mass to a somewhat lesser degree) without noticeably deteriorating the quality of the fit.

\begin{figure}[tb]
\begin{center}
    \includegraphics[width=0.5\textwidth]{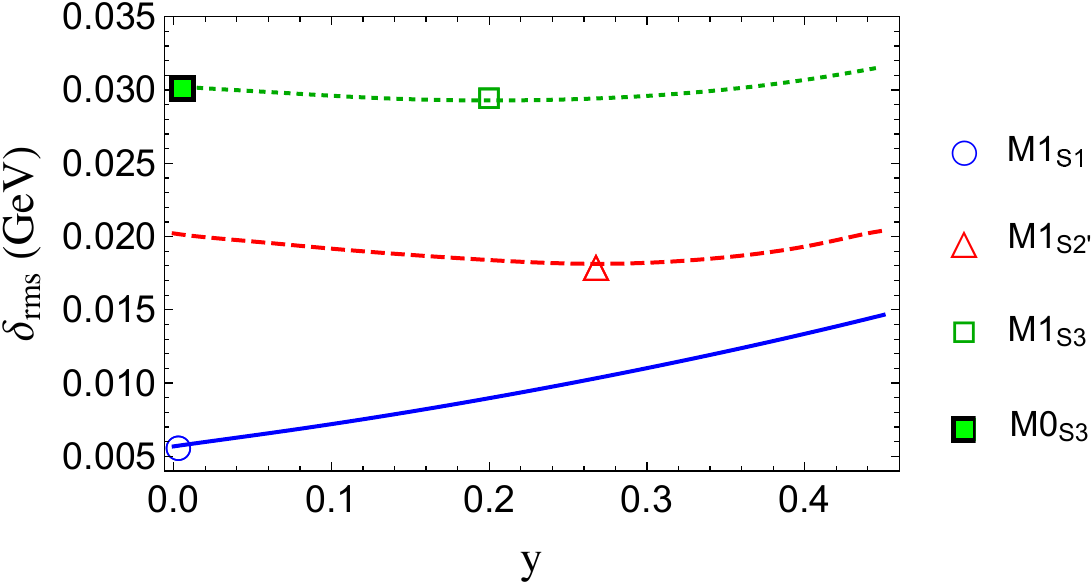}
    \vspace{-2mm}
\caption{Rms difference between calculated and experimental meson masses as a function of the mixing parameter $y$, using data set S1 (solid line), S2$'$ (dashed line), and S3 (dotted line). The symbols indicate the values for the models shown in Fig.~\ref{fig:spectrum}.}
\label{fig:rms-y}
\end{center}
\end{figure}

The Lorentz mixing parameter still prefers $y=0$ when we fit to pseudoscalar states only (set S1), but a minimum between $y=0.2$ and $0.3$ is found when larger data sets are used. To see if this contribution of V coupling is significant, we perform a series of calculations where the value of $y$ is fixed at different values and all other parameters are refitted. 
Figure~\ref{fig:rms-y} shows the rms difference to various data sets as functions of $y$. Clearly, the minima for data set S2 and S3 are so shallow that values of $y$ between 0 and roughly $0.25$ (for S3) are equally compatible with the meson spectrum. 

\begin{figure}[bt]
\begin{center}
    \vspace{-8mm}
    \includegraphics[width=0.5\textwidth]{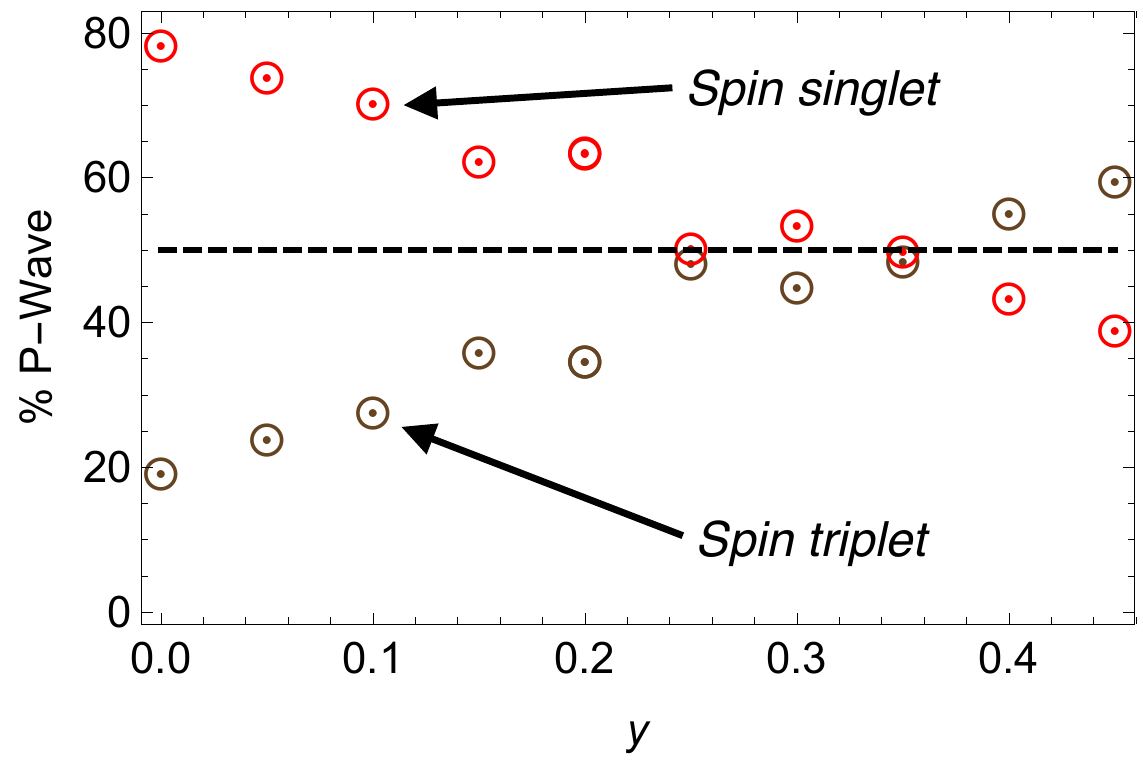}
    \vspace{-2mm}
\caption{Probabilities of spin triplet and singlet $P$ waves in the charmonium ground-state as a function of the mixing parameter $y$.}
\label{fig:cc-prob}
\end{center}
\end{figure}

We can therefore not make a definite conclusion about whether the confining interaction is of pure S+P nature, or if it contains also V coupling, by looking at the meson masses only. 
However, other observables may be more sensitive to $y$. As an example, Fig.~\ref{fig:cc-prob} shows sizeable changes in the contributions of spin singlet and triplet $P$ waves to the norm integral of the $c\bar{c}$ ground-state wave function when $y$ is varied in the series of fits described above. Although wave functions are not observables themselves, there are observables that depend strongly on the detailed structure of the wave functions, such as decay constants. By studying the dependence of decay constants on $y$ we should be able to obtain more precise constraints on the Lorentz structure of the confining interaction. This work is planned for the near future.

We thank the Funda\c c\~ao para a Ci\^encia e a Tecnologia (FCT) for support under contracts SFRH/BD/92637/2013, SFRH/BPD/\-100578/\-2014, and \linebreak UID/FIS/0777/2013.
\vspace{-3mm}

\end{document}